# Warnings and Caveats in Brain Controllability

Chengyi Tu[1,7], Rodrigo P. Rocha[1,2,7], Maurizio Corbetta[3,4,7], Sandro Zampieri[5,7], Marzo Zorzi[6,7,8] & S. Suweis[1,7*]

[1]*Dipartimento di Fisica e Astronomia, 'G. Galilei' & INFN, Università di Padova, Padova, IT.* [2]*Departamento de Física, Universidade Federal de Santa Catarina, 88040-900, Florianópolis-SC, Brazil.* [3]*Dipartimento di Neuroscienze, Università di Padova, Padova, IT.* [4]*Departments of Neurology, Radiology, Neuroscience, and Bioengineering, Washington University, School of Medicine, St. Louis, USA.* [5]*Dipartimento di Ingegneria dell'informazione, Università di Padova, Padova, IT.* [6]*Dipartimento di Psicologia Generale, Università di Padova, Padova, IT.* [7]*Padua Neuroscience Center, Università di Padova, Padova, IT.* [8]*IRCCS San Camillo Hospital Foundation, Venice, IT,*

There is large consensus that the complex, self-organizing structure of the human brain can be well described by the mathematical framework based on network theory. A recent article by Gu et al.[1] proposed to characterize brain networks in terms of their "controllability", drawing on concepts and methods of control theory. The analysis of controllability has the potential to unveil how specific nodes and/or sets of nodes control the dynamics of the entire network[1-4] and thus might provide insights on whether manipulating the local activity of specific nodes would fully or partially restore network functions after brain damage. Gu et al. applied the tools of control theory to quantify how the networks structure, defined by human connectome data (i.e., white matter pathways derived from diffusion tensor or diffusion weighted imaging), constrain or facilitate changes in brain state trajectories. They proposed that single brain regions that are densely connected facilitate the brain to easily reach many of its cognitive states (i.e. high average controllability), while weakly connected nodes promote the movement of the brain to difficult-to-reach states (i.e. high modal controllability). In other words, they claim that brain activity is controllable by a single node, and topology of brain networks provides an explanation for the types of control roles that different regions play in the brain. Here we present new results based on the analysis of four dataset and numerical simulations that challenge their main conclusion and undermine their use of the controllability framework.

In order to study single node controllability (in Kalman sensu), the authors first normalize the matrix describing the brain structural connectivity, and then calculate the controllability Gramian from each node (see Methods). Their first finding is that the brain can be *theoretically controllable* by a single region/node, i.e. the smallest (in absolute value) of the eigenvalues of the controllability Gramian $\lambda_{min}$ (see Methods) from each brain region as a control node is greater than zero. Then they measure the average controllability and the modal controllability from each node (see Methods), and find that average controllability is strongly correlated with the weighted degree, and the modal controllability is strongly anti-correlated with the weighted degree. Finally they evaluate the control contribution of different brain regions associated with known cognitive process (i.e. resting state networks (RSNs)) finding that different RSNs have different control roles.

To investigate further their results, we apply the controllability framework on four empirical open access datasets of large-scale structural brain connectivity from different studies[5–8]. In contrast to Gu et al.[1], we observe that for all the analyzed brain networks $\lambda_{min}$ is not greater than zero (we note that already in their work, also Gu et al.[1] have found that the smallest eigenvalue of their controllability Gramian from each brain region as a control node is $\lambda_{min} = 2.5 \times 10^{-23} \pm 4.8 \times 10^{-23}$ and thus it is *not* significantly greater than zero). Indeed, according to our data, the controllability Gramian from each node is not invertible. Therefore, it is not possible to assess the global controllability of the brain performed by a single region/node. We highlight that the structural properties of the DTI/DSI network data (symmetric with no self-loops) crucially affect the controllability of the system. Moreover, it is important also to quantify the energy needed to control the system from a specific region. In fact, although Gu et al.[1] have found that on average $\lambda_{min} \sim 10^{-23}$, in order to control such a system from a single region (node) would need an energy $\varepsilon \approx \lambda_{min}^{-1} \sim 10^{23}$, in practice the system is not controllable. Our results are in accordance with prior theoretical results[9] showing that for symmetric networks (like DTI/DSI brain connectivity data) the energy ε needed to control from a single node the system grows exponentially with the network size (i.e. to practically control the system (ε not too large) a fixed fraction of the networks nodes is necessary[9]).

Next, we study the role of nodes topological properties in controlling the brain network. Inspired by Gu and collaborators claim on the importance of nodes centrality, we first rank all networks nodes according to five distinct centrality measures. Then we build two ranking lists, following respectively the ascending and descending order of centralities magnitude. Then we generate a list of nodes in random order, irrespective of the nodes properties. For each of these three nodes sequences, we look for the minimum subsets of nodes needed to control the brain network. We have repeated this experiment for all brain connectivity datasets as well as for synthetic brain networks generated with a desired network topology (see Methods and Table 1). If topology is important we expect that the minimum number of control nodes will depend on the specific nodes properties (e.g. the most or the least central nodes). Our results are presented in Table 1. It shows the size of the minimum subset of control nodes with respect to five different centralities measures for the APOE-4 data[5] ($N = 110$ nodes), random Barabasi-Albert (BA) scale free networks[10] and uniform networks[10], both with the same number of nodes, edges, and edge weight distribution. No matter which centrality measure is adopted and how we generated the node sequences, the size of the minimum subset is always greater than one and similar in all three cases (see Table 1). This result strongly suggests that the topological structure of the DTI/DSI connectivity networks do not play an important role in brain controllability. We thus want to better understand the relationships between average/modal controllability and weighted degree and test if it is due to the specific brain network topology, or rather if it holds also for random networks obtained by randomizing the structure of real brain data, indeed, even if the controllability Gramian is not invertible, the average/modal controllability can still be calculated. Following Gu et al. (Fig. 2 b and d in1), we plot the average of the ranks for weighted degrees versus average of the ranks of the average controllability in Fig. 1a, and the average of ranks for weighted degrees versus the average of ranks of the modal controllability in Fig. 1c, (for details on average rank plots see Gu et al.'s methods section) of our four dataset. As Fig. 1a and 1c show, we find the same relation between controllability and network centrality reported by Gu et al. We then randomize the structural connectivity data and calculate average/modal controllability as a function of nodes properties in the random networks. Surprisingly, as shown in Fig. 1b and 1d, we found exactly the same relationship as before. We stress that the relationships in average controllability and modal controllability between real and randomized networks are not an artifact of the rank-rank plots. In fact, controllability measures between real and randomized networks are not distinguishable also by examining unranked magnitudes.

Finally we want to test if different RSNs are involved in different modes of controllability. To do that, we have built RSNs for the Hagmann dataset (based on RSNs template in the literature[11]). Following Gu et al. procedure, we consider the thirty nodes with highest average or modal controllability, and then calculate a vector $c_{data}$, where its $j$-th component gives the percentage of the nodes that belongs to the $j$-th RSN. We then repeat the same analysis for a proper random null model[10,12], obtaining $c_{rand}$ from random networks with the same degree distribution of the original dataset, but with no spatial correlations among regions (as observed in real RSNs). We find that both real and randomized data display very similar results on the modes of controllability of the different RSNs; indeed the Pearson correlation between $c_{data}$ and $c_{rand}$ is $\rho = 0.7$ (P-value $\approx$ 0.05). In other words we find that there are not specific roles played by RSNs in controlling the brain.

In summary, our results challenge the main conclusions of Gu et al work on brain controllability. Using the same methods and analyses on four datasets we find that the minimum set of nodes to control brain networks is always larger than one. We also find that the relationships between the average/modal controllability and weighted degrees also hold for randomized data and the there are not specific roles played by RSNs in controlling the brain. In conclusion, we show that there is no evidence that topology plays specific and unique roles in the controllability of brain networks. Accordingly, Gu et al. interpretation of their results[1, 13–15], in particular in terms of translational applications (e.g. using single node controllability properties to define target region(s) for neurostimulation) should be revisited. Though theoretically intriguing, our understanding of the relationship between controllability and structural brain network remains elusive.

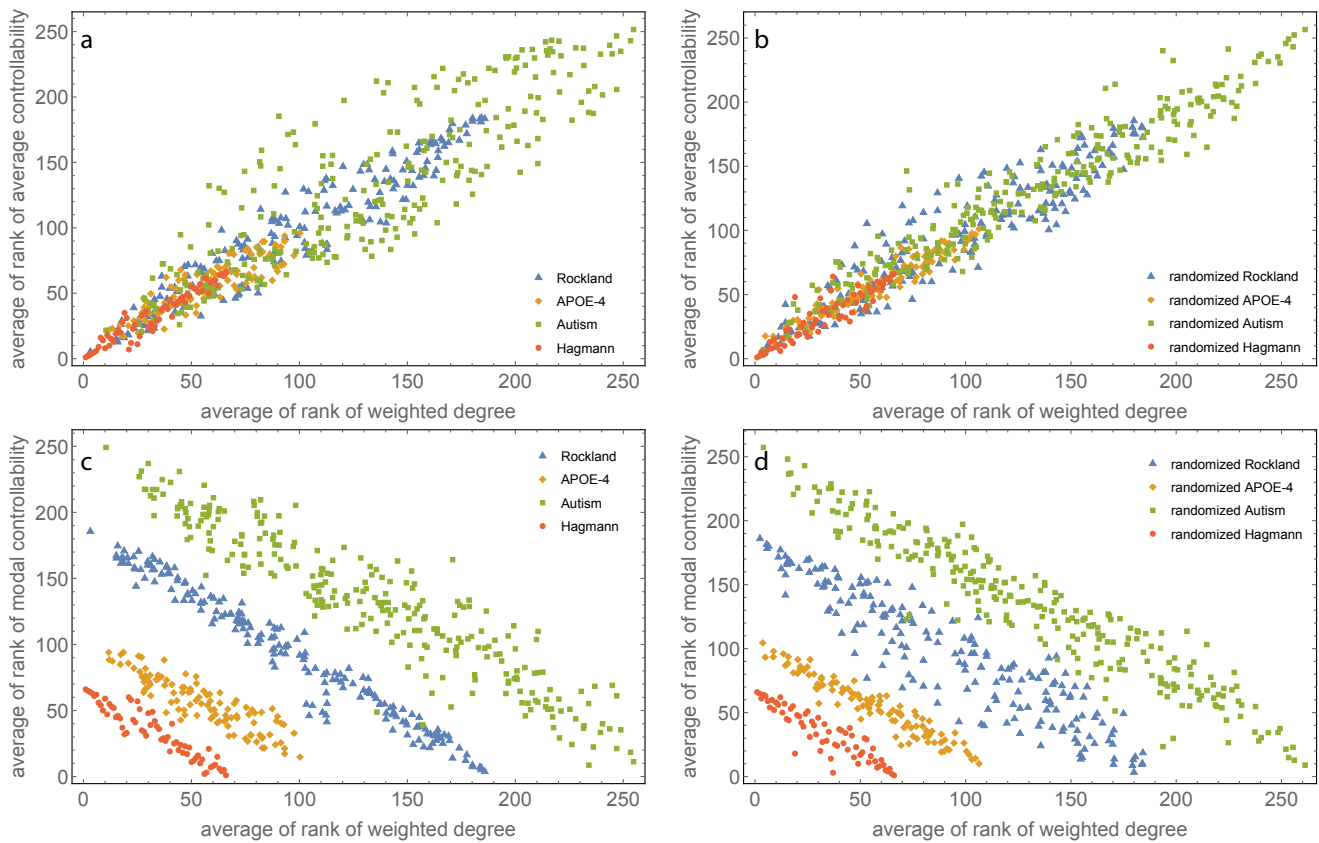

**Figure 1:** Comparing controllability measures between empirical data and randomized data. Scatter plot of the average of the ranks for weighted degrees versus average of the ranks of the average controllability for (a) the empirical data and (b) their randomized counterpart. Scatter plot of the average of ranks for weighted degrees versus the average of ranks of the modal controllability for (c) empirical data and (d) their randomized counterpart.

|  | APOE-4 data | | BA network | | Uniform network | |
| --- | --- | --- | --- | --- | --- | --- |
| Centrality measure | Low | High | Low | High | Low | High |
| Degree centrality | 31/0.28/86.16 | 30/0.27/83.37 | 28.04/0.25/77.95 | 28.04/0.25/77.91 | 28.16/0.26/78.27 | 27.8/0.25/77.27 |
| Betweenness centrality | 29/0.26/80.60 | 31/0.28/86.15 | 28/0.25/77.84 | 28.16/0.26/78.25 | 27.68/0.25/76.94 | 27.96/0.25/77.71 |
| Eigenvector centrality | 30/0.27/83.34 | 31/0.28/86.15 | 28.04/0.25/77.95 | 27.8/0.25/77.25 | 28/0.25/77.83 | 27.84/0.25/77.38 |
| Pagerank centrality | 29/0.26/80.60 | 30/0.27/83.37 | 28/0.25/77.84 | 27.88/0.25/77.47 | 28.16/0.26/78.27 | 27.8/0.25/77.27 |
| Random sequence | 30/0.27/83.38 | 30/0.27/83.37 | 27.04/0.25/77.94 | 28.04/0.25/77.93 | 27.46/0.25/76.33 | 27.66/0.25/76.88 |

**Table 1:** The minimum number / fraction (with respect the size of the network) of nodes and its energy Trace{W$_K$} to guarantee the controllability of the APOE-4 data and the corresponding BA and uniform random networks (of same size, connectivity and edge weight distribution) following the procedure described in the main text. For a given sequence, "High" is the rank from the largest to the smallest, while "Low" is the rank from the smallest to the largest. Hagmann, Autism and Rockland dataset have been also analyzed and we found qualitatively the same results, i.e. in order to control the network we need a fraction of nodes ranging from 16% to 25% of the whole brain, regardless of the centrality measures used to rank the nodes.

## Methods

**Neural dynamics model.** Similarly to Gu et. al., we employed a simplified linear discrete-time neural dynamics model. This model is a noise-free variation of the Galán model[2] and can be derived from the linearization of a general Wilson-Cowan system (for strengths and weakness of this model see Refs. (2),(3),(12)). We then discretized the linearized dynamics and obtained the following equation:

$$x(t + \Delta t) = Ax(t) + B_K u_K(t) \qquad (1)$$

where $x(t)$ is the state variable representing the neural activity of the brain regions, $A = (1 - \alpha\Delta t)I + cM\Delta t$ is the $N \times N$ Jacobian matrix, $\alpha$ is the inverse of the relaxation time, $\Delta t$ is the discretized time step, $I$ is the $N \times N$ identity matrix, $M$ is the symmetric and weighted DTI structural matrix and c is a normalization constant. Following the original work of Galan[2], we fixed $\alpha = 1.0, \Delta t = 0.2$. The input matrix $B_K$ identifies the set of control points $K$ in the brain, where $K = \{k_1, \cdots, k_m\}$ and

$$B_K = [e_{k_1}, \cdots, e_{k_m}], \qquad (2)$$

where $e_i$ denotes the $i$-th canonical vector of dimension $N$. The input $u_k : R_{\geq 0} \to R^m$ denotes the control strategy.

**Controllability Gramian.** Given the neural dynamics, Eq. (1), we can set up the Lyapunov equation, $W_K - AW_K A^T = B_K B_K^T$. If A is stable, then this equation has a unique solution called Gramian:

$$W_K = \sum_{m=0}^{\infty} A^m B_K B_K^T (A^T)^m \qquad (3)$$

For the discretized dynamics (1) and $A$ symmetric, stability is ensured if all eigenvalues of $A$ lies in the interval $\lambda_i \in (-1,1)$ con $i=1,\ldots,N$. Therefore, in order to guarantee stability of our dynamical system we set the normalization constant to $c = 1/(1 + \lambda_{\max})$, where $\lambda_{max} = Max\{|\lambda_i|, i = 1,\ldots, N\}$. Following this procedure we constrain the eigenvalue spectrum of $A$ to be in the desired interval.

The results presented in the main text were obtained as follows. In Figure 1, we applied the controllability framework in each of the individual DTI matrices within a given dataset, $M_i$, where $i$ ranges from 1 to $n$, the total number of healthy subjects. We calculate the average and the modal controllability for each of the corresponding Jacobian matrix $A_i$ and then we performed the average of the rank plots following the procedure adopted by Gu et. al. The results presented in Table 1 were employed using the averaged DTI matrix, i.e. $M_{av} = \sum M_i/n$.

**Controllability measures.** The average and the modal controllability measures employed in this study follows the expressions reported by Gu et. al.[1]. The average controllability can be approximate by Trace$\{W_K\}$. As shown by Gu and collaborators[1], the strong correlation between node degree and average controllability can be understood analytically and it does not depend on the specific topological properties of the structural connectivity M. The modal controllability represents the ability of a node to control each evolutionary mode of a dynamical network. It is given by $\phi_i = \sum_{j=1}^{N}\left(1 - \lambda_j(A)\right)^2 v_{ij}^2$ where $V = v_{ij}^2$ is the eigenvector matrix and $(\lambda_1,\cdots,\lambda_N)$ are the corresponding eigenvalues of $A$.

**Synthetic random networks.** We generated two type of synthetic DTI random networks, namely, the Barabasi-Albert scale-free networks[10] and the uniform networks[10], using the same number of nodes, edges and edge weight distribution of the averaged DTI matrix, $M_{av} = \sum M_i/n$ for each dataset. We implemented these constraints by fitting the edge weights of $M_{av}$ to a Pareto distribution[10]. The synthetic networks were generated using standard routines available in the Mathematica software. Controllability measures displayed in Table 1 were averaged over 50 realizations of the synthetic random networks.

**Centralities measures.** Degree centrality refers to the number of links a node has to other nodes[10]. Betweenness centrality is a measure of centrality based on shortest paths and is given by $\sum_{s\neq v\neq t} n_{st}(v)/n_{st}$ where $n_{st}$ is the total number of shortest paths from node $s$ to node $t$ and $n_{st}(v)$ is the number of those paths that pass through the node $v$[10]. Eigenvector centrality is based on the centralities of its neighbor nodes and can be expressed as $c = \frac{1}{\lambda} A^T c$, where $A$ is the adjacency matrix and $\lambda$ is the corresponding largest eigenvalue[10]. PageRank centrality is a generalization of the eingevector centrality, weighting in a non-linear way the number and quality of links connected to the node[10].

**DTI brain connectivity datasets.** We applied the controllability framework on four empirical datasets of large scale structural (DTI or DSI) brain connectivity from different studies [5–8]. All datasets are open access and were obtained from the Human Connectome Project[4], namely, the APOE-4 dataset[5] ($n = 30$ APOE-4 non-carrier and n=25 APOE-4 carrier individuals; gray matter parcellation into $N = 110$ large scale regions); The Rockland dataset[6] ($n = 195$ healthy subjects, $N = 188$ large scale regions); The Hagmann dataset[8] (average matrix corresponding to $n = 5$ healthy subjects, $N = 66$ cortical regions) and the Autism dataset[7] ($n = 94$ healthy subjects, $N = 264$ large scale regions). For more specific details on the data acquisition and preprocessing we refer to the original studies.

**Randomized data set.** In Figure 1 we randomized the topological structure of the DTI matrix $M$ by randomly rewiring all its matrix elements, but keeping symmetry. The randomized DTI matrices are thus symmetric random with zero diagonal entries. In building random RSNs, we randomized the DTI matrix $M$ by randomly rewiring all its matrix elements, but fixed the degree sequence of the original dataset[12].

**References.**

**Authors Contributions:** SS., S.Z, M.Z., M.C. designed the research, C.T., R.R., S.S. performed the research, R.R., C.T. analyzed the data. All authors contributed in writing the manuscript.